\documentclass[prb,twocolumn]{revtex4-1}

\usepackage{amsmath} 
\usepackage{graphicx} 
\usepackage{subfigure}

\begin{document}

\title{Particle sliding on a turntable in the presence of frictional forces}
\author{Akshat Agha}
\affiliation{BITS Pilani, K. K. Birla Goa Campus, Goa, India}
\author{Sahil Gupta}
\affiliation{BITS Pilani, K. K. Birla Goa Campus, Goa, India}
\author{Toby Joseph}
\email{Electronic mail: toby@goa.bits-pilani.ac.in}
\affiliation{BITS Pilani, K. K. Birla Goa Campus, Goa, India}
\date{26th June, 2014}

\begin{abstract}
Motion of a point particle sliding on a turntable is studied. The equations of motion are derived assuming that the table exerts 
frictional force on the particle, which is of constant magnitude and directed opposite to the direction of motion of the particle relative 
to the turntable.  After expressing the equations in terms of dimensionless variables, some of the general properties of the
solutions  are discussed. Approximate analytic solutions are found for the cases in which (i) the particle is released from
rest with respect to the lab frame and, (ii) the particle is released from rest with respect to the turntable. 
The equations are then solved numerically to get a more complete understanding of the motion. It is found that
one can define an escape speed for the particle which is the minimum speed required to get the particle to move off to
infinity. The escape speed is a function of the distance from the center of the turntable and for a given distance from the center, 
it depends on the direction of initial velocity. A qualitative explanation of this behavior is given in terms 
of the fictitious forces. Numerical study also indicates an alternative way for measuring the coefficient of friction between the
particle and the turntable.
\end{abstract}

\maketitle

\section{Introduction} Study of particle motion in rotating frames is a crucial part of an undergraduate mechanics
course. Earth is a rotating frame and hence the study of any large scale motion on the surface of the Earth as observed by an Earth bound observer would 
necessitate the use of such a rotating frame. The use of a rotating frame may also simplify the study of certain mechanics problems that involves 
rotating bodies in the laboratory. One of the first examples discussed in an undergraduate lecture on rotating frames is that of the motion of a particle on
a rotating two dimensional platform. The turntable is used extensively in setting up demonstration experiments
to help students understand the fictitious Coriolis and centrifugal forces \cite{bligh, hume, levine}. A rotating platform on which
a particle is moving  also forms the background to a large number of text book problems elucidating Coriolis and centrifugal forces
and the application of the polar coordinate system. 

There are a few problems that are exactly solvable in the context of motion of a particle on a turntable. 
The easiest to work out is the motion of the particle that is not coupled to the table (sliding without friction). 
In that case, the motion is one with uniform velocity when observed from the lab frame. This linear motion when observed from the 
turntable is not so simple and one has to invoke Coriolis and centrifugal forces to explain the resulting trajectory\cite{bush,lenka}. 
A more difficult problem that can be exactly solved is that of a spherical ball moving without slipping on a turntable \cite{burns}. 
The motion of the ball in this case is similar to that of the motion of a charged particle in a magnetic field. 
The ball follows circular trajectories whose location and radius varies according to the initial conditions. The variations 
of this problem where the turntable is tilted \cite{burns,romer} and freely spinning \cite{weckesser} have been studied. 
Solution to the first problem yields cycloidal motion and for the second  case the trajectories are conic sections. 
Introducing sliding effects renders the problem difficult to treat analytically. The motion of a ball that rolls with sliding on a turntable 
has been studied using computer simulations and experiments \cite{ehrlich}. 
 
Though most of the problems related to turntables in physics textbooks deal with friction, many of them study the motion only in a limited context. 
For example, a typical problem would involve finding the maximum angular frequency with which the table can rotate if the particle is to remain at rest 
with respect to the table when the particle is placed at a given distance from the axis of rotation. But as we shall see, the system offers an interesting set of behaviors 
if one is willing to probe a little further. A detailed study of the general motion under such circumstances is something that is missing in the literature. 
The present work is an attempt to fill that gap by studying the motion of a point particle on a turntable in the presence of dry friction, both analytically 
and numerically.

In the following section, the equations of motions for the sliding particle are derived and expressed in dimensionless form. Some of the general
properties of the solutions to the equations are then discussed. In section III, the solution to the equations are derived for two special cases: (i) when the particles is
moving slowly with respect to the lab frame and (ii) when the particle is moving slowly with respect to the turntable. 
Section IV deals with the numerical solution of the problem. Various kinds of trajectories that are possible are illustrated and the 
dependence of particle escape speed on the initial direction of motion is determined at various locations on the table. 
In the final section we conclude with a brief summary and possible extensions of the current work.

\section{The equations of motion}
Consider a point particle of mass $m$ on a turntable of infinite extent. Let the coefficient of friction
between the particle and the table surface be $\mu$ (we will assume that the coefficients of static and kinetic friction
are the same). Assume the table is rotating with a uniform angular speed, $\Omega$, about an axis that is perpendicular to the plane
of the table. Let $(r, \theta)$ be the polar coordinates of the particle, defined by taking the point of intersection of the rotation axis and
the table as the origin. For a particle at rest with respect to the table, the frictional force on the particle will be in the radial direction. 
The magnitude of this static friction force will vary from $0$ to $\mu m g$, depending on the distance of the particle from the origin. 
When there is relative motion between the particle and the table, the magnitude of kinetic friction force on the particle will be $\mu m g$. 
The direction of frictional force in this case is determined by the velocity of the particle relative to the table, 
\begin{equation}
\vec v_{rel} =  (\dot r \hat  r + 
r \dot \theta \hat \theta) - r \Omega \hat \theta .
\end{equation}
Since the frictional force opposes the relative motion between the two, its direction will be opposite to that of $\vec v_{rel}$.
The frictional the force on the particle, when it is in motion with respect to the table, is given by
\begin{equation}
\vec F = \mu m g (-\hat v_{rel}) = \mu m g \frac{r \Omega \hat \theta - (\dot r \hat r 
+ r \dot \theta \hat \theta)}{\sqrt{r^2(\Omega - \dot \theta)^2 + {\dot r}^2}},
\label{thequ}
\end{equation}
where $\hat v_{rel}$ is the unit vector in the direction of the relative velocity.
The equations of motion in polar coordinates take the form,
\begin{equation}
\ddot r - r {\dot \theta}^2 = \frac{-\mu g \dot r}
{\sqrt{r^2(\Omega - \dot \theta)^2 + {\dot r}^2}}
\label{eqmn1}
\end{equation}
and
\begin{equation}
2 \dot r \dot \theta + r {\ddot \theta} = \frac{\mu g r(\Omega - \dot \theta)}
{\sqrt{r^2(\Omega - \dot \theta)^2 + {\dot r}^2}}
\label{eqmn2}
\end{equation}
As mentioned above, if the particle remains at rest with respect to the table (that is, $\dot r = 0$ and
$\dot \theta = \Omega$) the force on the particle will be radial and its magnitude will be $F_s = m \Omega^2 R$, where 
$r = R$ is the radial location of the particle.

The time scale in the problem is set by the angular velocity of the turntable,
$\Omega$. There is also an inherent length scale in the problem, $R_{max}$, which is the
distance beyond which the particle cannot remain at rest with respect to the turntable.
This distance is found by equating the centrifugal force $m \Omega^2 R_{max}$ in the rotating frame at $r = R_{max}$,
to the maximum value $\mu m g$ of the static friction force. This gives
\begin{equation}
R_{max} = \frac{\mu g}{\Omega^2}.
\end{equation}
Define two dimensionless variables $\tau \equiv t \Omega$ and $\xi \equiv {r}/{R_{max}}$.
Equations (\ref{eqmn1}) and (\ref{eqmn2}) in terms of these new variables are
\begin{equation}
\xi'' - \xi {\theta'}^2 = \frac{-\xi'}
{\sqrt{\xi^2(1 - \theta')^2 + {\xi'}^2}}
\label{finfeq1}
\end{equation}
and
\begin{equation}
2 \xi' \theta' + \xi {\theta''} = \frac{\xi (1 - \theta')}
{\sqrt{\xi^2(1 - \theta')^2 + {\xi'}^2}}
\label{finfeq2}
\end{equation}
where the prime denotes differentiation with respect to the variable $\tau$.

The equations obtained above are two coupled nonlinear equations and a complete analytic solution looks difficult. 
The non-conservative nature of the frictional force makes it impossible to write down the first integrals
of motion using conservation laws. Nevertheless, one can discern certain general properties of the 
solution:
\begin{enumerate}
\item If the particle is at rest with respect to the table within the
region given by $\xi < 1$, it remains at rest with respect to the table.
In rotating frame of the turntable, particle velocity is zero initially and hence the Coriolis force
will be absent at that initial instant. The centrifugal force acting in the radial direction will be balanced by the static friction
force because $r < R_{max}$. In particular, if the particle comes to rest with respect to the table in the internal region given 
by $\xi < 1$, the particle will remain at rest with respect to the table subsequently.
\item From Eq. (\ref{finfeq1}), since $\xi \ge 0$, if $\xi' \le 0$, then $\xi'' \ge 0$.
This implies that if initially $\xi' > 0$ (that is with the radial
component of the velocity pointing outwards), then $\xi$ will increase monotonically with $\tau$. This is because
if $\xi'$ becomes zero, it can only become positive the next instant (since $\xi''$ would be larger than or
equal to zero). If initially $\xi' < 0$ (radial velocity is pointing inwards), then $\xi$ will reach a minimum value 
and then will increase monotonically with time. However, in both cases, if $\theta'$ takes the value $1$ when $\xi'$ turns $0$, 
and if this happens for a $\xi < 1$, the particle will come to rest in the frame of the table.
\item From Eq. (\ref{finfeq2}), it is seen that if $\theta' = 0$, then $\theta'' > 0$. This implies that, if one starts with $\theta' > 0$  
(that would mean a tangential velocity in the direction of the rotation of the turntable), it will remain so at all future times. 
\item From the analysis above it is seen that $\xi'$ eventually becomes larger than or equal to $0$. After long enough times
(long enough to ensure $\xi'$ is positive) if $\theta' \ge 1$, then from Eq. (\ref{finfeq2}) 
it is seen that $\theta'' < 0$. Also under similar conditions if $\theta' < 0$, then $\theta'' > 0$ (as can be seen from Eq. (\ref{finfeq2})). 
The above two conditions would ensure that eventually the value of $\theta'$ lies between $0$ and $1$.
\end{enumerate}

The conclusions that we arrived at above are what one expects intuitively. If the particle does not come to rest with respect to the
table, it will eventually move away from the center to greater distances. And its angular velocity, as observed from the lab frame, will 
eventually lie between $0$ and angular velocity of the table, $\Omega$.

\section{Analysis for special cases}
Two cases where one can approximate the equations of motion to a form in which
analytical solutions are possible are discussed below. These are when: (A) the particle is moving slowly when observed from
the lab frame and, (B) the particles is moving slowly when observed from the frame of the turntable.

\subsection{Particle moving slowly in the lab frame}
\begin{figure}
\begin{center}
\includegraphics[width=9cm]{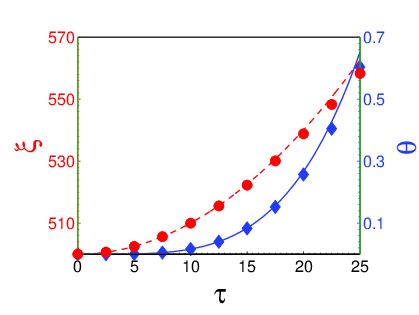} 
\caption{Comparison of approximate analytic solution with numerical simulation for the case when the particle is released from rest with respect to the lab frame.
The particle is released at the location, $\xi_0 = 500$. The theoretical curves for $\xi(\tau)$ (dashed curve) and $\theta(\tau)$ (solid curve)
gives good match with the results from the simulation (circles for $\xi$ and diamonds for $\theta$). Note that $\tau$ is
time measured in units of of $\frac{1}{\Omega}$ and $\xi$ is distance measured in units of $R_{max}$.}
\label{fig01}
\end{center}
\end{figure}
Imagine that the particle has been placed gently on to the turntable by a person who is at rest in the lab frame. This would mean that the particle
velocity is zero initially and it will remain small for a certain period of time at least. The particle velocity can be considered to be small if 
$|\dot r \hat r + r \dot \theta \hat \theta| \ll \Omega r$. That is, $\dot r \ll \Omega r$ and $\dot \theta \ll \Omega$. 
In terms of the dimensionless variables, the conditions becomes $\xi' \ll \xi$ and $\theta' \ll 1$. If the initial condition is such that $\xi'(0) \ll \xi(0)$ and 
$\theta'(0) \ll 1$, there will be a time range in which $\xi'$ and $\theta'$ remain small compared to $\xi$ and $1$ respectively. 
During that time range, $\theta'$ can be neglected compared to $1$ in the equations of motion. The equations of motion, 
Eq. (\ref{finfeq1}) and Eq. (\ref{finfeq2}), then reduce to
\begin{equation}
\xi'' = \xi {\theta'}^2 - \frac{\xi'}
{\sqrt{\xi^2 + {\xi'}^2}}
\end{equation}
and
\begin{equation}
{\theta''} = \frac{1}
{\sqrt{\xi^2 + {\xi'}^2}} - \frac{2 \xi' \theta'}{\xi}
\end{equation}
To simplify further, neglect $\xi'$ in comparison to $\xi$ and replace $\xi$ with $\xi(0) (\equiv \xi_0)$, the value of $\xi$ at $\tau= 0$. 
This means the resultant equations will be valid until the value of $\xi'$, which is initially small, has not grown in size to be 
comparable to $\xi$ and $\xi$ itself has not changed appreciably from its initial value. These approximations will have to be taken 
into account when estimating range of validity of the analysis (see below). 
Under these assumptions, the radial equation is 
\begin{equation}
\xi''  = \xi_0 {\theta'}^2 - \frac{\xi'}
{\xi_0}
\end{equation}
and the tangential equation becomes
\begin{equation}
{\theta''} = \frac{1}{\xi_0}.
\label{tangeqn}
\end{equation}
Integration of Eq. (\ref{tangeqn}) with  initial conditions taken to be $\theta(0) = 0,\; \theta'(0) = 0$ gives,
\begin{equation}
\theta = \frac{p^2}{2 \xi_0}.
\end{equation}
The choice $\theta(0) = 0$ can always be made by choosing the direction to the initial location
of the particle from the center of the table as the reference direction for measuring $\theta$.
The solution for $\theta$  substituted into radial equation gives a second order inhomogeneous equation
for $\xi$. The solution with initial conditions $\xi(0) = \xi_0$ and $\xi'(0) = 0$ gives,  
\begin{equation}
\xi = \frac{p^4}{12 \xi_0} + \xi_0.
\end{equation} 

The approximations made in arriving at the above solution are $\theta' \ll 1, \;\xi' \ll \xi$ and  
$|\xi - \xi_0| \ll \xi_0$. For the solution obtained these conditions reduce to $\tau \ll \xi_0$, $\tau \ll \xi_0^{2/3}$
and $\tau \ll  \sqrt{\xi_0}$ respectively. The last of the above conditions is the one that gets violated first 
as time progresses and this happens for times below $\tau \sim \sqrt{\xi_0}$.  In Fig. \ref{fig01} the trajectories obtained by
numerical integration of the exact equations of motion are compared to the approximate analytic solutions. The numerical work was
carried out by programming in MATLAB and will be discussed in greater detail in the section on numerical results. 
The value of $\xi_0$ used was $500$ and the particle is at rest in the lab frame at $\tau = 0$. The agreement between the
numerical and analytical solutions is found to be good. The solution should be valid till $\tau \sim 23$ and this is approximately where
the curves start deviating from the numerical results. 

\subsection{The particle moving slowly with respect to the turntable}
Consider the case where the particle is kept at some location on the turntable with zero initial velocity with respect to the turntable.  
The particle velocity with respect to the turntable will then be small at least for some interval of time.
Making the relevant approximations one can solve for the motion of the particle that would be valid during this interval. 
When the particle is moving slowly with respect to the turntable $\theta' \approx 1$ and $\xi' \approx 0$. 
Define 
\begin{equation}
\Theta' \equiv \theta'-1,
\end{equation}
which is the angular velocity of the particle in the frame fixed to the turntable. The condition that $\theta' \approx 1$ implies that
magnitude of $\Theta' \ll 1$.
Neglecting the term $\xi \Theta'$ as compared to $\xi$ in the radial equation and $2\xi'\Theta'$ as compared to $2 \xi'$ in the tangential
equation, the equations of motion become
\begin{equation}
\xi'' - \xi = \frac{-\xi'}
{\sqrt{\xi^2 \Theta'^2+ {\xi'}^2}}
\end{equation}
and
\begin{equation}
2 \xi' + \xi {\Theta''} = -\frac{\xi \Theta'}
{\sqrt{\xi^2 \Theta'^2 + {\xi'}^2}}\;.
\end{equation}
Since $\xi' = \Theta' = 0$ initially, the term $\xi^2 \Theta'^2$ can be discarded in comparison to $\xi'^2$
inside the square root in both the equations. The justification for this approximation is that, it is $\xi'$ that grows faster with time
than $\xi \Theta'$. As the particle starts from rest in the frame of the turntable, the velocity independent centrifugal force tends to increase the radial velocity
whereas the tangential velocity build up can happen only once the velocity dependent Coriolis force is appreciable. The two equations, under the above
approximation becomes
\begin{equation}
\xi'' - \xi = -1
\label{rad2}
\end{equation}
and
\begin{equation}
2 \xi' + \xi {\Theta''} =0,
\label{th2}
\end{equation}
where in the second equation, the term $\frac{\xi \Theta'}{\xi'}$ has been neglected. The resulting equation 
for the radial motion [Eq. (\ref{rad2})] is identical to the equation for radial motion of a bead on a uniformly rotating rod in the presence of friction. 
The solution to this equation is
\begin{equation}
\xi = (\xi_0 - 1) \cosh \tau+ 1
\label{xi2}
\end{equation}
where the boundary conditions $\xi(0) = \xi_0$ and $\xi'(0) = 0$ have been used to determine the constants of integration.
Note that the above solution is valid only when $\xi_0 > 1$. If $\xi_0 < 1$, the particle will remain 
at rest with respect to the table. In the limit of $\xi_0$ approaching $1$, the equation predicts the values of $\xi = 1$ for all times, as expected.

\begin{figure}
\begin{center}
\includegraphics[width=8cm]{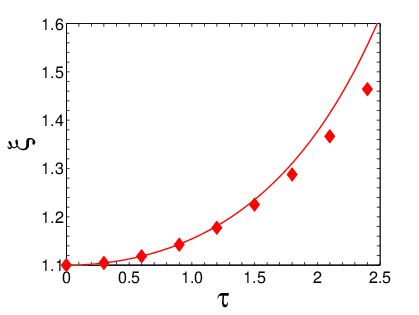} 
\caption{Comparison of theoretical analysis of $\xi(\tau)$ with numerical results for the case when the particle is released at
the location, $\xi_0 = 1.1$ with zero relative velocity with respect to the turntable. The theoretical curve (solid line)
matchs the results from the simulation (diamonds) for times for which $\tau \lesssim 1$}
\label{fig02}
\end{center}
\end{figure}

Integrating Eq. (\ref{th2}) once and using the boundary conditions above,
\begin{equation}
\Theta' = 2 \log(\frac{\xi_0}{\xi}).
\end{equation}
The solution obtained for $\xi$ can be substituted in above equation and integrated to get the variation of $\Theta$ 
with time. The above solutions is valid provided $|\Theta'| \ll 1$ and $|\xi \Theta'| \ll |\xi'|$. For values of $\xi_0 \gg 1$, 
these conditions are satisfied for $\tau \ll 1$. For values of $\xi_0$ closer to (and greater than) $1$, the approximations hold for longer times. 
A comparison of the solutions from the above approximate analysis and the numerical results is give in Fig. \ref{fig02}.  

\section{Numerical analysis of the equations}
The analysis carried out above is limited in its scope and it does not yield the rich variety of 
trajectories that are possible in the system. In this section a numerical solution to the exact equations is presented and
the resulting trajectories are classified in terms of their long time behavior. The equations of motion (see Eq. (\ref{thequ})) 
expressed in the Cartesian coordinate system are, 
\begin{eqnarray}
m\dot v_x&=&  \mu m g \frac{-\Omega \; y- v_x}{\sqrt{(\Omega \; y + v_x)^2 + (\Omega \; x - v_y)^2}} \nonumber \\
m\dot v_y &=& \mu m g \frac{\Omega \; x - v_y}{\sqrt{(\Omega \; y + v_x)^2 + (\Omega \; x - v_y)^2}} \nonumber \\
\dot x &=& v_x \nonumber \\
\dot y &=& v_y \nonumber \\
\end{eqnarray}
The four coupled differential equations above have been solved using the second order Runge-Kutta method for a variety of initial conditions. The unit for time
is chosen such that the magnitude of $\Omega$ is $1$ and distances are measured in units of $R_{max}$. This allows comparison to be made between 
the numerical results and the analytical results above (see Figs. \ref{fig01} and \ref{fig02}). The time step, $\delta t$, in the simulation was $0.001$.
This was found to be sufficient for convergence of the trajectories to acceptable limits for the entire time duration for which trajectories were generated.

\begin{figure}
\begin{center}
\includegraphics[width=8cm]{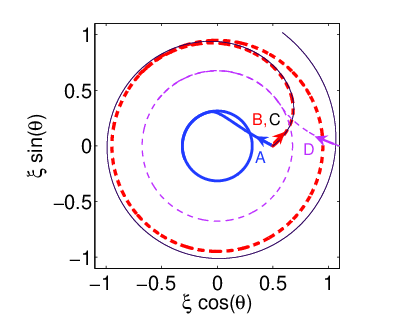} 
\caption{Trajectories of the particle as seen from the lab frame for various initial conditions. The parameters varied are $\xi_0$ (initial location), $\phi$ 
(the direction of initial velocity measured in the turntable's frame with respect to the radially outward direction) and $v_0$ (the initial speed). The arrows indicate
the starting location and direction of initial velocity. The initial 
conditions for various trajectories are: $\xi_0 = 0.5$, $\phi = \pi$ and $v_0 = 1$ for trajectory A (red, solid line), $\xi_0 = 0.5$, $\phi = 0$ and $v_0 = 0.6000$ 
for trajectory B (red, dashed line), $\xi_0 = 0.5$, $\phi = 0$ and $v_0 = 0.6078$ for trajectory C (black, dotted line) and, $\xi_0 = 1.1$, $\phi = 1.2 \pi$ and 
$v_0 = 1.3$ for trajectory D (pink, dash-dot line.} 
\label{fig03}
\end{center}
\end{figure}
\begin{figure}
\begin{center}
\includegraphics[width=8cm]{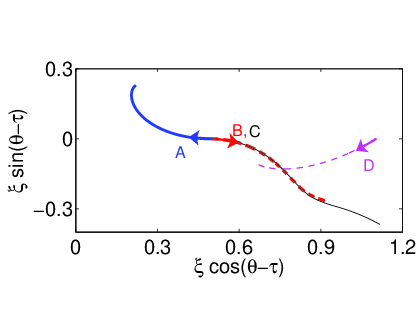} 
\caption{Trajectories of the particle as seen by an observer on the turntable. The parameters for the trajectories A, B, C and D are the same as that in
Fig. \ref{fig03}. The arrows indicate the starting location and direction of the initial velocity. In all the cases except C (black, dotted line) the particle 
has come to rest with respect to the table.}
\label{fig04}
\end{center}
\end{figure}

Some of the trajectories (as seen from the lab frame) obtained from the simulations are shown in Fig. \ref{fig03}. The same trajectories as seen from the rotating frame
are given in Fig. \ref{fig04}. One can see that the particle trajectories exhibit a variety of behavior. The trajectory depends on the point at which the 
particle is released, the speed with which it is released, and the direction of the initial velocity. In trajectories A and B (see figure) the particle starts off in the region $\xi < 1$ 
and comes to rest with respect to the table. In trajectory C, which has almost the same initial conditions as trajectory B but for a marginal increase in the initial speed, 
the particle manages to come out of the $\xi = 1$ region and eventually moves off to infinity. Trajectory D shows another possibility where the particle is projected from the 
region $\xi > 1$ but the initial conditions are such that it comes to rest with respect to the table in the region $\xi < 1$. One can classify these trajectories into those 
executing bounded motion and those that move off to infinity. As discussed above (section II), the bound trajectories all correspond to the particle coming 
to rest with respect to the turntable in the region $\xi < 1$.
\begin{figure}
\begin{center}
\includegraphics[width=8cm]{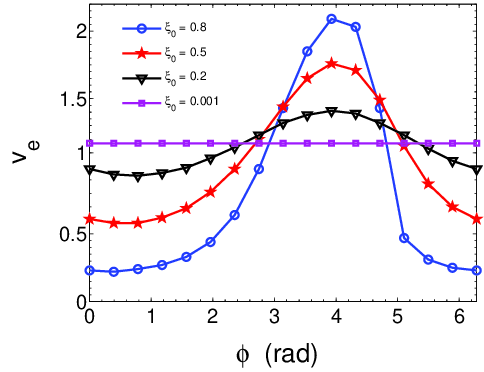}
\caption{The plot shows the escape speed, $v_e$, as a function of $\phi$. Both the speed and the angle are given as seen by the person located at the point of release
of the particle on the turntable and at rest with respect to the turntable. The angle $\phi = 0$ corresponds to the radial direction. The plots are for the release of particle at 
$\xi_0 = 0.8$ (circles), $\xi_0 = 0.5$ (triangles),  $\xi_0 = 0.2$ (inverted triangles) and $\xi_0 = 0.001$ (squares). The escape speed has a non-monotonic 
behavior with $\phi$.}
\label{fig05} 
\end{center}
\end{figure}

It is interesting to note that the notion of an escape speed, defined as the minimum initial speed required to get the particle to move off to infinity, 
exists for the particle moving on a turntable. From the simulations, the escape speed is found to depend on the location where the 
particle is released as well as the direction of its initial velocity. This is unlike the the case for motion in a conservative central force field where 
escape speed is independent of the direction of velocity. The variation of the escape speed, $v_e$ (defined with respect to the observer on the turntable) 
as a function of direction of initial velocity, $\phi$ (as measured by an observer on the turntable with respect to the radially outward direction), were determined 
from the simulations for various initial values of position, $\xi_0$. The results are shown in Fig. \ref{fig05} for the cases when $\xi_0 < 1$. 
For a given $\xi_0$, the escape speed is found to vary with $\phi$. The minima of the escape speed occurs at $\phi \approx 0.4$ radians for all values of $\xi_0$.
Qualitatively this makes sense, because for a noninertial observer on the turntable the Coriolis force tries to deflect particle to the right of its direction of 
motion and it should be advantageous to release the particle to the left of the radially outward direction (rather than the radial direction itself) so as to make it move 
farther away from the center. For the same reason, it is most difficult to get the particle to move away to infinity if it is released at an angle of about $\phi \approx 4$ radians 
with respect to the outward radial direction. In this case, the Coriolis force tries to push the particle towards the center of the table confining it more to the interior region.
The variation of escape speed with $\phi$ is more complicated when the particle is located outside the $\xi < 1$ region. From the definition of $\xi$ it is 
clear that a particle left at rest with respect to the table at location $\xi > 1$ will move off to infinity. But it is possible to get the particle to come rest in the region $\xi < 1$
for a range of initial speeds that depend on the angle $\phi$ (trajectory D in Fig. \ref{fig03}, for example). 

\begin{figure}
\begin{center}
\includegraphics[width=8cm]{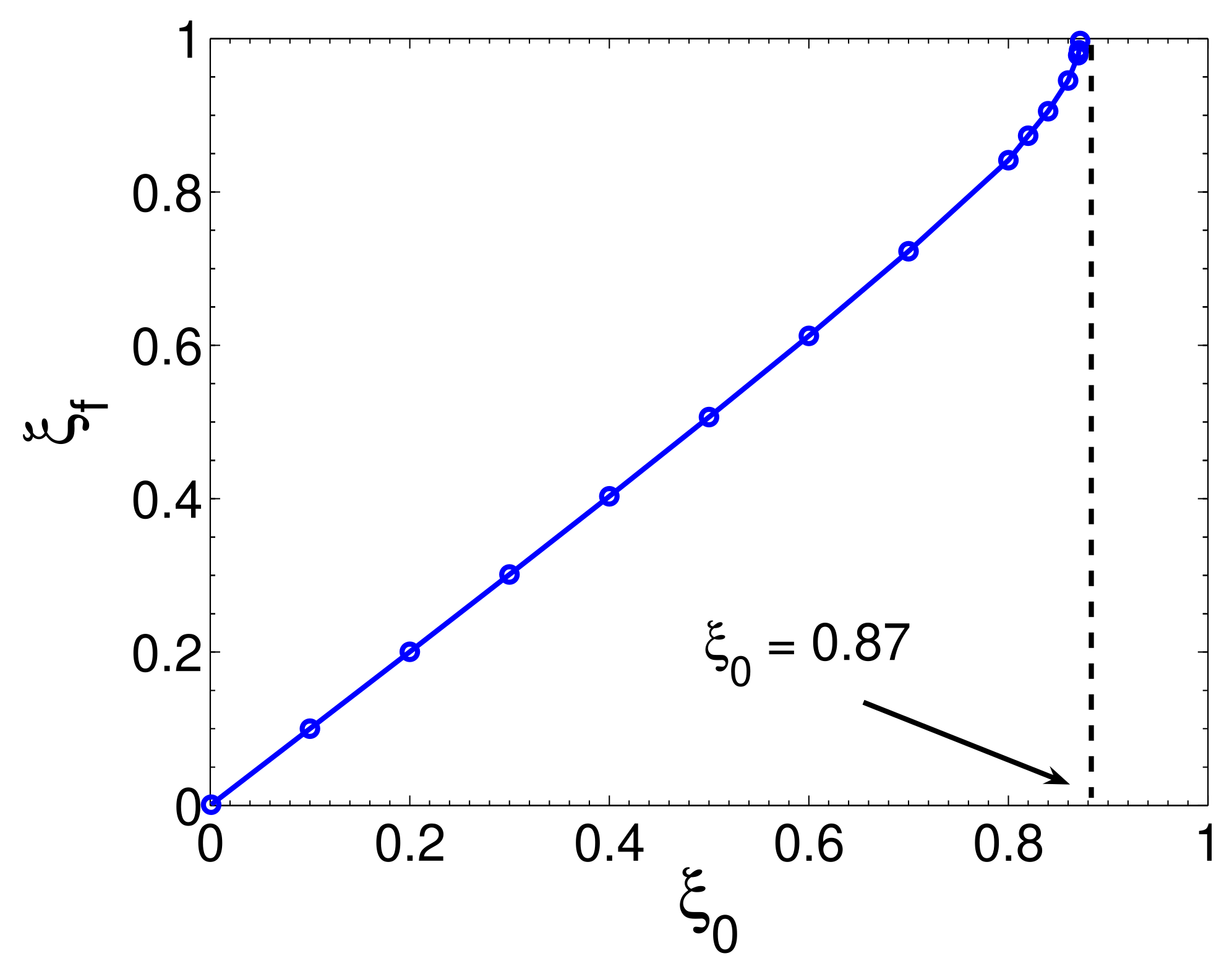} 
\caption{The final location of the particle as a function of the initial distance from the center for the cases when particle is released at rest with 
respect to the lab frame. For initial location $\xi_0 > 0.87$, the particle moves off to infinity.}
\label{fig06}
\end{center}
\end{figure}

What is the final position of the particle that is placed gently on the table by the inertial observer? In this case the particle is
initially at rest in the lab frame. The result obtained form the simulations is shown in Fig. \ref{fig06}, which gives the final location of the
particle as a function of the initial distance from the center for distances such that the particle comes to rest with respect to the turntable. 
The maximum distance at which the particle can be placed and it will still remain 
bounded is $\xi_{0c} = 0.87$. Experimental determination of this distance could be a way of finding the frictional coefficient between the particle and the table. If $r_{0c}$
is the actual distance corresponding to $\xi_{0c}$ value, then
\begin{equation}
\mu = \frac{\Omega^2 r_{0c}}{\xi_{0c} \;g}.
\end{equation} 
This method is to be contrasted with the more familiar one wherein one would place the particle at a known location, $r_a$, with table at rest initially and find 
the minimum rotational speed of the turntable at which the particle starts to slide. If $\Omega_{min}$ is the maximum value of the angular speed for a distance $r_a$, then
\begin{equation}
\mu = \frac{\Omega_{min}^2 r_a}{g}.
\end{equation} 
Though the expression for $\mu$ looks similar in both the above equations, the underlying concept employed is subtly different.

\section{Summary}
The problem of a particle sliding on a rotating table in the presence of friction has been studied both numerically and analytically.
The system offers an interesting set of behavior that is overlooked in a typical text book treatment of this problem. The equations may be expressed 
in a dimensionless form using the inherent length and time scales in the problem for a systematic analysis. The motions of the particle for short durations 
after it has been released from rest (both with respect to the turntable and with respect to the lab) may be found analytically. 
Numerical integration of the equations have been carried out using second order Runge-Kutta algorithm 
for a variety of initial conditions. Interestingly, the notion of an escape speed is well defined for this system and it is a function of location 
of the particle and the direction of initial velocity. Our analysis also suggests an alternative way of determining the coefficient of friction between 
the particle and the table.
 
All the results we have discussed in the paper should be verifiable in an undergraduate lab with a turntable set up. It has to be kept in mind that the analysis we have carried
out above assumes that the particle is point like. Effect of finite size of the particle could be important as is the case for example, in the motion of a puck subject to
friction on a stationary surface \cite{voyenli,denny}. There are two types of effects that can be important when one considers the
finiteness of the particle concerned. One is the tendency of the object to topple and the other is the tendency to rotate about an
axis perpendicular to the plane of the turntable and passing through the center of mass. The tendency to topple can be mitigated by having an object that has
height, $h$, much less than its lateral dimension, $L$ and the second type of rotation can be neglected, provided the particle is located 
at distances large compared to its lateral size, $L$ (see Appendix). Thus a sufficiently small object on a rotating table should show a behavior similar to the 
predictions made in this paper.  

A natural extension to the work presented here would be to study the motion 
that results when the coupling between the particle and the table are not of the dry friction type. Preliminary analysis of Stokes' drag coupling indicates a completely 
different kind of dynamics.

\appendix*
\section{}
\begin{figure}
\begin{center}
\includegraphics[width=4cm]{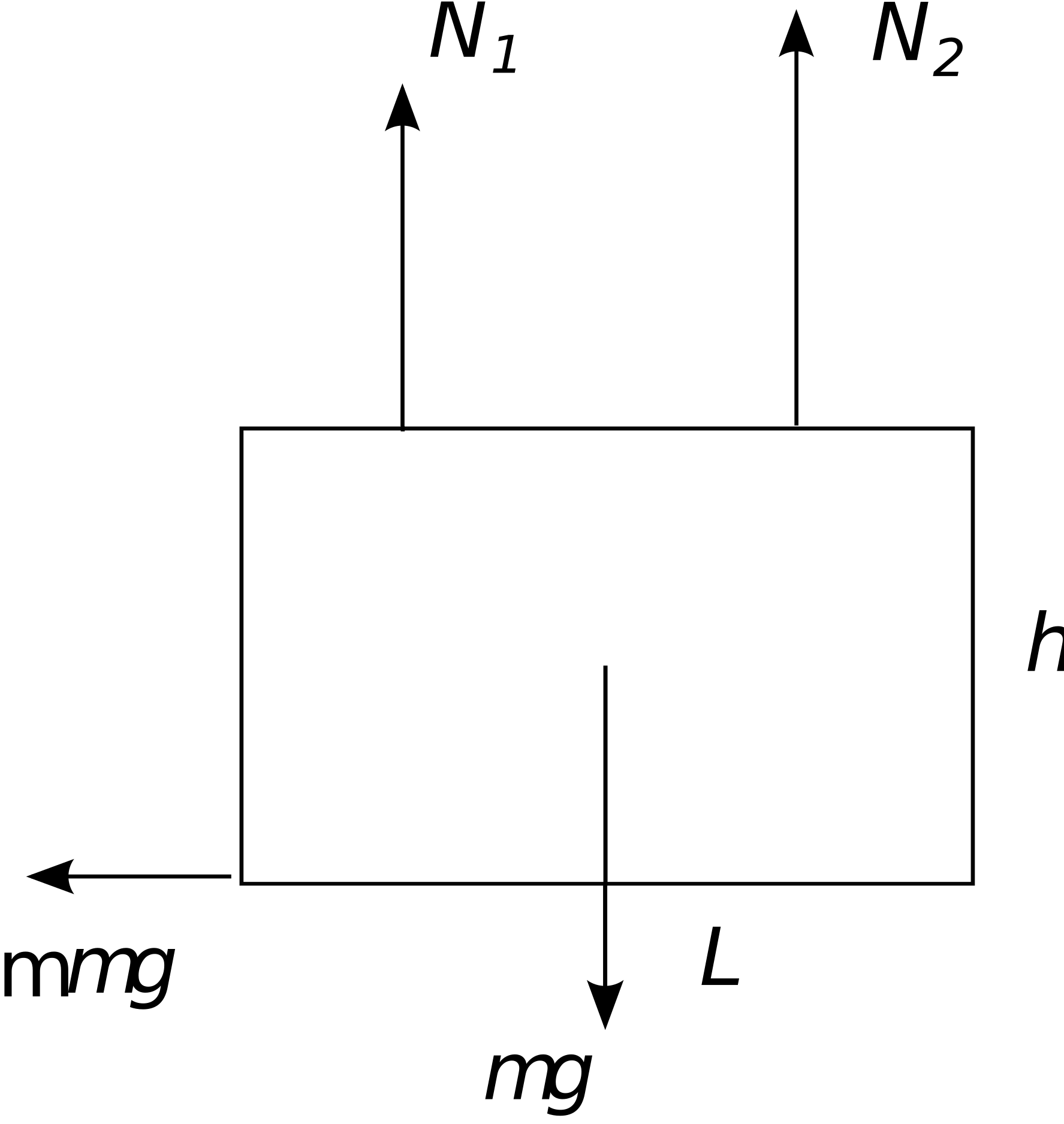} 
\caption{Free body diagram for a body of height $h$, lateral size $L$ and mass $m$, purely translating on the turn table. Various forces acting on
the body are indicated.}
\label{fig07}
\end{center}
\end{figure}
In the calculations above the assumption of point particle has been made. An order of magnitude analysis of the conditions required for this assumption to be
valid are derived here. 

The tendency for the particle to topple is considered first. Consider a particle that has a lateral size, $L$, and height $h$, the exact 
shape being immaterial for the approximate analysis carried out. Assuming that the particle is undergoing pure translational motion, the forces acting
on the particle are the force of gravity, the normal reaction force and the frictional force pointing in the direction opposite to the direction of the relative velocity
of the particle with respect to the table. To find when the particle would topple we need to compute the component of the torque about the center of mass that 
lies in the plane of the table. To make the analysis simple, it is assumed that the reaction forces are acting at two points on the surface of contact that are located 
at distance $L/4$ on either side of the center of mass along the direction of the relative velocity of the body with respect to the turntable (see Fig. \ref{fig07}).
The condition for rotational equilibrium about an axis passing through the center of mass and parallel to the surface is
\begin{equation}
\mu m g \frac{h}{2} + N_1 \frac{L}{4} - N_2 \frac{L}{4} = 0.
\end{equation}
The condition for non translation of the center of mass in the vertical direction gives
\begin{equation}
N_1 + N_2 = mg.
\end{equation}
Solving the above equations one gets,
\begin{equation}
N_2 = \frac{mg}{2} + \mu mg \frac{h}{L}.
\end{equation}
Requiring $N_1 \approx N_2$ (otherwise parts of the body will tend to loose contact with the table) gives the condition $\mu \frac{h}{L} \ll 1$. Since $\mu$ is
typically of order $1$ or less, the above condition will be satisfied provided $h \ll L$.

\begin{figure}
\begin{center}
\includegraphics[width=4cm]{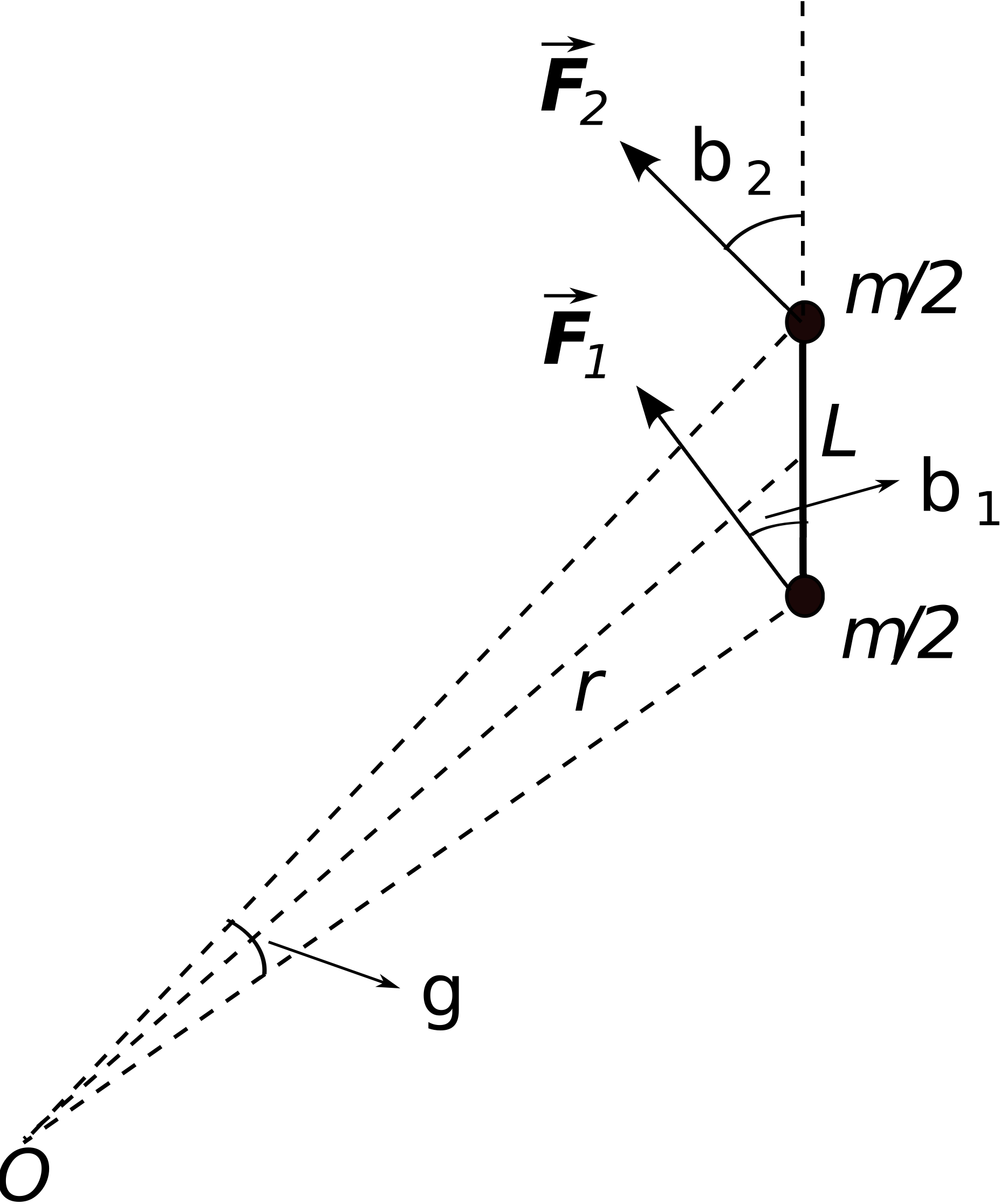} 
\caption{A rigid body consisting of two point masses of mass $m/2$ each and separated by a massless rigid rod of length $L$. The body is located at a distance
$r$ from the center of the turntable and it subtends an angle $\gamma$ at the center. The $\vec F_1$ and $\vec F_2$ are the frictional forces acting on the 
masses and they make angles $\beta_1$ and $\beta_2$ respectively with the rod connecting the masses.}
\label{fig08}
\end{center}
\end{figure}
The tendency to rotate about an axis parallel to the direction of $\Omega$ (that is, rotation in the plane of the turntable) 
can happen if there is a net torque in the vertical direction. Even if the body were purely translating, it is possible for various parts of the body to have 
relative velocity with respect to the turntable that points in different directions due to the fact that it has finite size. As a result of this, the frictional force 
acting at different regions of the contact area could be pointing in different directions leading to a non zero torque about the center of mass in the vertical direction. 
This can in turn lead to an in-plane rotation of the particle even if the particle has pure translational motion to start with. If the linear velocity of the parts of the object due to 
this rotation becomes comparable to the translational velocity of the center of mass, the equations of motion that were used to study the motion of the particle 
will not yield accurate results. 

To get an order of magnitude estimate of when the finiteness of the particle will modify the
equation of motion for the center of mass (Eq. (\ref{thequ})), consider a body consisting of two point particles of mass $m/2$ each,
rigidly attached with a massless rod of length $L$ (see Fig. \ref{fig08}) and located at a distance $r$ from the center of the turntable. 
Assuming the body to be at rest with respect to the lab frame, the torque on the body is given by
\begin{equation}
\tau = \mu \frac{m g}{2} \frac{L}{2}(\sin \beta_2 - \sin \beta_1),
\end{equation}
where the different angles are indicated in Fig. \ref{fig08}. But $\beta_2 = \beta_1 + \gamma$ and $\gamma$ is of the order $L/r$, provided $L \ll r$.
Substituting and simplifying one gets
\begin{equation}
\tau \lesssim \frac{\mu m g L^2}{r}
\end{equation}
where  $\cos \gamma \approx 1$, $\sin \gamma \approx \frac {L}{r}$ and $\cos \beta_1 \lesssim 1$ have been used to get order of magnitude value.
The angular acceleration, $\alpha$, produced by this torque is obtained by dividing the torque by the moment of inertia about the center of mass,
$I = m L^2/2$.  This gives,
\begin{equation}
\alpha \lesssim \frac{\mu g}{r}.
\end{equation}
The linear acceleration of the masses corresponding to the angular acceleration $\alpha$ is 
\begin{equation}
a \sim {\mu g}\frac{L}{r}.
\end{equation} 
On the other hand the net force acting on the center of mass is of the order $\mu m g$ (assuming $\cos \gamma \approx 1$) and the corresponding linear
acceleration of the center of mass is,
\begin{equation}
a' \sim \mu g.
\end{equation}
Comparing $a$ and $a'$, it is seen that effects of rotation of the body can be neglected provided $L \ll r$ for an object that was initially at rest in the lab frame. 

Thus the two conditions required for the body to be treated as a point particle are that its lateral size be much larger than its height and that 
it is location from the center of the table is at distances large compared to its lateral size.
The conditions derived above for the validity of treating the body as a point particle is only approximate. For example, one could have an initial condition
in which rotational motion is of such a magnitude that the frictional force acting at different locations cannot be treated as being collinear. Nevertheless,
this analysis gives us a ballpark figure of when the results derived in the paper are valid.

\begin{acknowledgments}
We thank Gaurav Dar for careful reading of the manuscript and suggesting numerous corrections. We also thank the two referees for their
suggestions to improve the content and presentation. This article may be downloaded for personal use only. Any other use requires prior permission 
of the author and AIP Publishing. This article appeared in ( Am. J. Phys, 83, 126 (2015)) and may be found at (https://doi.org/10.1119/1.4896664)
\end{acknowledgments}

\end{document}